\documentclass[useAMS,usenatbib]{mn2e}
\usepackage{epsfig}
\input psfig.sty
\usepackage{graphicx}% Include figure files
\usepackage{dcolumn}% Align table columns on decimal point
\usepackage{bm}% bold math
\usepackage{epsfig}
\usepackage{graphicx}% Include figure files
\usepackage{dcolumn}% Align table columns on decimal point
\usepackage{bm}% bold math
\newcommand{\bq}{\begin{equation}}		%%
\newcommand{\eq}{\end{equation}}		%%
\newcommand{\bqn}{\begin{eqnarray}}		%%
\newcommand{\eqn}{\end{eqnarray}}		%%
\newcommand{\lb}{\label}			%%
\begin{document}

\title[Testing the cosmic distance duality with X-ray gas mass fraction and supernovae data]{Testing the cosmic distance duality with X-ray gas mass fraction and supernovae data}
\author[R. S. Gon\c{c}alves, R. F. L. Holanda and J. S. Alcaniz]{R. S. Gon\c{c}alves\thanks{E-mail: rsousa@on.br }, R. F. L. Holanda\thanks{E-mail: holanda@on.br } and J. S. Alcaniz\thanks{E-mail: alcaniz@on.br } \\
Observat\'orio Nacional, 20921-400, Rio de Janeiro - RJ, Brasil\\
\pagerange{\pageref{firstpage}--\pageref{lastpage}} \pubyear{2010}}

\maketitle

\label{firstpage}

\begin{abstract}

In this letter we discuss a new cosmological model-independent test for the cosmic distance duality relation (CDDR), $\eta = D_{L}(L)(1+z)^{-2}/D_{A}(z)=1$, where $D_{A}(z)$ and $D_{L}(z)$ are the angular diameter and luminosity distances, respectively. Using the general expression for X-ray gas mass fraction ($f_{gas}$) of galaxy clusters, $f_{gas} \propto D_L{D_A}^{1/2}$, we show that $f_{gas}$ observations jointly with type Ia supernovae (SNe Ia) data furnish a validity test for the CDDR. To perform our analysis we use 38 $f_{gas}$ measurements recently studied by two groups considering different assumptions to describe the clusters (La Roque {\it{et al.}} 2006 and Ettori {\it{et al.}} 2009) and two subsamples of SNe Ia distance luminosity extracted from the Union2 compilation. In our test we consider the $\eta$ parameter as a function of the redshift parameterized by two different functional forms. It is found that the La Roque {\it{et al.}} (2006) sample is in perfect agreement with the duality relation  ($\eta = 1$) whereas the Ettori {\it{et al.}}  (2009) sample presents a significant conflict.

\end{abstract}


%\tableofcontents

\section{Introduction}

The so-called {reciprocity relation}, proved long ago by Etherington (1933), is a fundamental result for observational cosmology (see, e.g., Schneidder et al. 1992 and Peebles 1993 for different cosmological analyses in which the relation is directly or indirectly used). It states that if source and observer are in relative motion, solid angles subtended between them are related by geometrical invariants which involve the source redshift $z$ measured by the observer (see Ellis 1971; 2007 and references therein).

Etherington reciprocity law can be presented in various alternative ways, either in terms of solid angles or relating astronomical distances. Probably, its most useful version in the context of cosmology, sometimes referred to as { cosmic distance duality relation} (CDDR), relateing the {luminosity distance} $D_{\scriptstyle L}$ with the {angular diameter distance} $D_{\scriptstyle A}$ through the identity
\begin{equation}
\frac{D_{\scriptstyle L}}{D_{\scriptstyle A}}{(1+z)}^{-2}=\eta , \quad \quad \mbox{with $\eta = 1$}\;.
  \label{rec}
\end{equation}
This result is {theoretically} valid for {all} cosmological models based on Riemannian geometry, being independent either upon Einstein field equations or the nature of matter. It only requires conservation of photon number and sources and observers to be connected by null geodesics in a general Riemannian spacetime (see, e.g., Ellis 1971, 2007).  In reality, any consistent observational deviation from Eq.(\ref{rec}) would give rise to a cosmological crises (Ellis 2007) with a clear evidence of a new physics.

Ideally, the CDDR should be tested from observations of cosmological sources whose intrinsic luminosities and sizes are known. Thus, after measuring the source redshift, one can determine both $D_{\scriptstyle L}$ and $D_{\scriptstyle A}$ to test directly the relation. In recent papers, the validity of the {CDDR} has been discussed using $D_A$ measurements of galaxy clusters obtained from their X-ray surface brightness plus Sunyaev-Zeldovich (SZE) observations and luminosity distances of type Ia supernovae (SNe Ia) (see, e.g., De Bernardis {\it{et al.}} 2006; Holanda {\it{et al.}} 2010; 2011a; Li {\it{et al.}} 2011; Nair {\it{et al.}} 2011). In such analyses,  subsamples of galaxy clusters and SNe Ia were built so that the difference in redshift between objects in each sample is of the order of $10^{-3}$, thereby allowing a validity test of the duality relation (see also Khedekar \& Chakraborti (2011) for a new version of the so-called Tolman test (Tolman 1932) based on future observations of a redshifted 21 cm signal from disk galaxies).

Another variant of the CDDR test also discussed in the recent literature assumes a cosmological background suggested by a set of observations and check the validity of the CDDR in the context of some astrophysical effect. Examples of this approach are given by Bassett \& Kunz (2004) who used $D_{\scriptstyle L}$ measurements from type Ia supernova (SNe Ia) data and $D_{\scriptstyle A}$ estimates from FRIIb radio galaxies (Daly \& Djorgovski 2003) and ultra compact radio sources (Gurvitz 1999) observations in order to test the possibility of new physics by assuming the $\Lambda$CDM cenario. Uzan {\it{et al.}} (2004) showed that observations from Sunyaev-Zeldovich effect and X-ray surface brightness from galaxy clusters also provides a test for the distance duality relation. By assuming the concordance $\Lambda$CDM model and using angular distances from 18 galaxy clusters (Reese {\it{et al.}} 2002), they found $\eta = 0.91^{+ 0.04}_{-0.04}$ (1$\sigma$), a value only marginally consistent with the standard result ($\eta=1$). Recently, Avgoustidis {\it{et al.}} (2010) assumed an extended CDDR, given by $D_L=D_A(1+z)^{2+\epsilon}$ to constrain cosmic opacity and found $\epsilon=-0.04_{-0.07}^{+0.08}$ (2$\sigma$) from a combination of SNe Ia and the latest measurements of the Hubble expansion lying in the redshift range $0 < z < 2$ (Stern {\it{et al.}} 2010). More recently, Holanda {\it{et al.}} (2011b)  used the validity of the CDDR in the $\Lambda$CDM framework to constrain possible galaxy cluster morphologies.

In this letter, we propose a consistent model-independent test for the CDDR by using $D_A$ measurements extracted from gas mass fraction observations of  galaxy clusters and $D_L$ from current SNe Ia data. To perform our analysis, we use two samples of 38  gas mass fraction measurements  obtained from X-ray observations, as discussed by La Roque {\it{et al.}} (2006) and Ettori {\it{et al.}} (2009), plus two sub-samples of the SNe Ia taken from the Union2 compilation (Amanullah et al. 2010). The SNe Ia redshifts of each sub-sample were carefully chosen to coincide with the ones of the associated galaxy cluster sample ($\Delta z < 0.006$), thereby allowing a direct test of the CDDR. This method has a clear advantage on tests involving SNe Ia and $D_A$ of galaxy clusters from their X-ray plus SZE observations since the error bars in  gas mass fraction measurements  are considerably smaller than those obtained from X-ray/SZE technique. It is important mentioning, however, that both methods have a common weak point since they rely on astrophysical observations  with different systematics errors sources which may influence estimates of the CDDR parameter $\eta$.

\section{Gas mass fraction and the CDDR}

The gas mass fraction is defined by $f_{gas}=M_{gas}/M_{Total}$, where $M_{gas}$ is the mass of the intracluster medium gas and $M_{Total}$ is the total mass including barionic mass and dark matter. As is well known, the baryonic matter content of galaxy clusters is dominated by the X-ray emitting intracluster gas via predominantly thermal bremsstrahlung (see, e.g., Sarazin  (1988) for more details).

Following Sasaki (1996), the gas mass $ M_{gas} (<R) $ within a radius $R$  derived by X-ray observation can be written as

\begin{eqnarray}
M_{gas} (<R) &=& \left( \frac{3 \pi \hbar m_e c^2}{2 (1+X) e^6}
\right)^{1/2}  \left( \frac{3 m_e c^2}{2 \pi k_B T_e} \right)^{1/4}
m_H \nonumber\\
& & \mbox{\hspace{-2.5cm}} \times \frac{1}{[\overline{g_B}(T_e)]^{1/2}}
{r_c}^{3/2} \left
[ \frac{I_M (R/r_c, \beta)}{I_L^{1/2} (R/r_c, \beta)} \right] [L_X
(<R)]^{1/2}\;,
\end{eqnarray}
where $m_e$ and $m_H$ are the electron and hydrogen masses, respectively, $X$ is the hydrogen mass fraction, $T_e$ is the gas temperature, $\overline{g_B}(T_e)$ is the Gaunt factor,  $r_c$ stands for the core radius and
$$
I_M (y, \beta) \equiv \int_0^y (1+x^2)^{-3 \beta/2} x^2 dx\;,
$$
$$
I_L (y, \beta) \equiv \int_0^y (1+x^2)^{-3 \beta} x^2 dx\;.
$$
Note that $L_X, r_c$ and $R$ are not directly derived from observations, but depend on the adopted cosmological model, i.e.,
\begin{equation}
L_X (<R) = 4 \pi [D_L(z; \Omega_i H_0)]^2 f_X(<\theta),
\end{equation}
\begin{equation}
r_c = \theta_c D_A(z; \Omega_i, H_0),
\end{equation}
\begin{equation}
R = \theta D_A(z; \Omega_i, H_0),
\end{equation}
where $f_X(<\theta)$ is the total bolometric flux within the outer angular radius $\theta$ and $\theta_c$ is the angular core radius (see also Peebles (1993) for more details). In the above equations, $\Omega_i$ stands for the energy density parameters of the assumed cosmological scenario and $H_0$ is the current value of the expansion rate. From the above expressions, the gas mass can be written as
\begin{equation}
M_{gas} (<\theta) \propto D_L{D_A}^{3/2}(z;\Omega_i).
\end{equation}
On the other hand, the total mass  within a given radius $R$  is obtained by assuming that the intracluster gas is in hydrostatic equilibrium, i.e.,
\begin{equation}
M_{tot} (<R) = - \left. \frac{k_B T_e R}{G \mu m_H} \frac{d \ln
n_e(r)}{d \ln r} \right|_{r=R},
\end{equation}
which for the spherical $\beta$ model profile (Cavaliere \& Fusco-Fermiano 1976) provides
\begin{equation}
M_{tot} (<\theta) \propto D_A .
\end{equation}
Therefore, the gas mass fraction defined earlier is in its more general form given by
\begin{equation}
\label{final}
f_{gas} \equiv \frac{M_{gas}}{M_{tot}} \propto D_L{D_A}^{1/2}\;.
\end{equation}
It is interesting to note that in most of the analyses discussed in the literature the relation $f_{gas} \propto {D_A}^{3/2}$ is readily assumed, although its validity is justified only in the cases in which Eq. (\ref{rec}) (and all its underlying theoretical assumptions) is satisfied.

We will define the $f_{gas}$ model function as
\bqn
\lb{GasFrac}
f_{gas}(z) &=& N \left[\frac{D_L^{*}{D_{A}^{*1/2}}}{D_L{D_{A}^{1/2}}}\right]\;,
\eqn
{ where the normalization factor $N$ carries all the information about the matter content in the cluster, such as stellar mass fraction,  non-thermal pressure and the depletion parameter $b$, which indicates the amount of cosmic baryons that are thermalized within the cluster potential (see, e.g., Eq. (3) of Allen {\it{et al.}} 2008 for more details).} The asterisk denotes the corresponding quantities in the fiducial model used in the observations.

The term between brackets accounts for deviations in the geometry of the Universe from this model, which makes the analysis model-independent (see, e.g., Allen {\it{et al.}} 2002; 2004 and Lima {\it{et al.}} 2003 for more details on the case in which $\eta = 1$ is assumed). Since in our analyses we use data obtained in the $\Lambda$CDM context to which $\eta = 1$, Eq. (\ref{GasFrac}) must be rewritten as
\bqn
\lb{GasFrac3}
f_{gas}(z) &=& N \left[\frac{{D_{A}^{*}}^{3/2}}{\eta{D_{A}}^{3/2}}\right],
\eqn
in such a way that the angular diameter distance that will be used in our analyses is given by
\bqn
\lb{GasFrac4}
D_A(z) &=& N^{2/3} \left[\frac{{D_{A}^{*}}}{\eta^{2/3}{f_{gas}}^{2/3}}\right].
\eqn
In what follows, we briefly discuss the data samples used in our statistical analysis.

\begin{figure*}
\label{Fig}
\centerline{\epsfig{figure=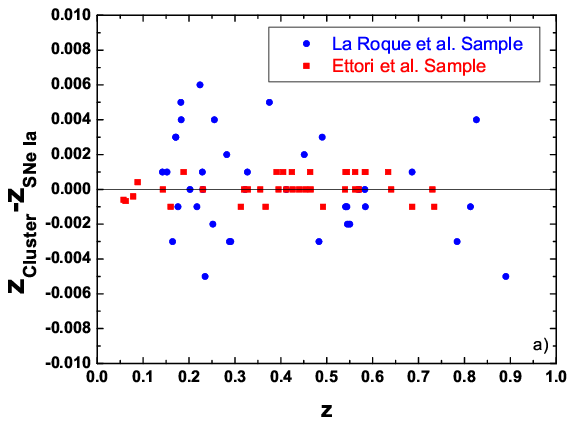,width=2.2truein,height=2.5truein}
\epsfig{figure=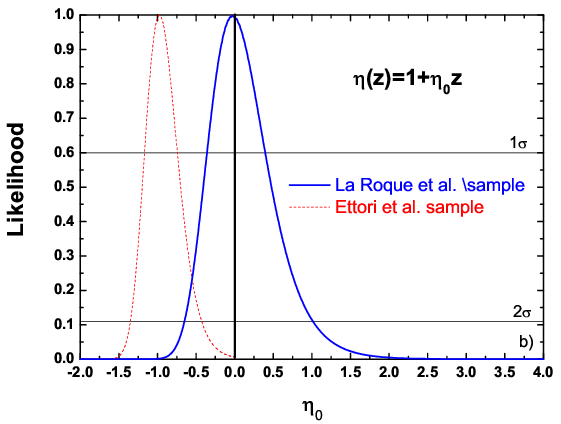,width=2.2truein,height=2.5truein}
\epsfig{figure=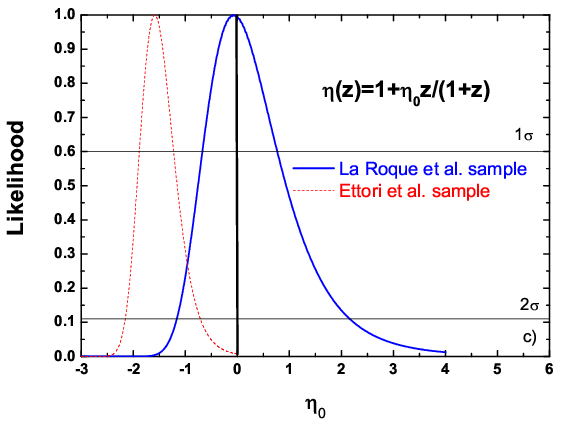,width=2.2truein,height=2.5truein}
\hskip 0.1in}
\caption{{{a)}} The redshift difference for each pair of SNe Ia/cluster from Union2 sub-samples and galaxy clusters of the La Roque {\it{et al.}} (2006) (filled blue circles) and the Ettori {\it{et al.}} (2009) (filled red squares) samples. {{b)}} The likelihood distribution functions for La Roque {\it{et al.}} (solid blue line) and Etorri {\it{et al.}} (red dashed line) samples using the linear parametrization P1. {{c)}} The same as in Figure 2 using the nonlinear parametrization P2. Comparing these results with those displayed in the previous figure, we clearly see that the $\eta$ parameterizations adopted in our analysis do not alter considerably the best-fit results.}
\end{figure*}

\section{Data sets}

\subsection{Gas mass fraction samples}

In order to perform our CDDR test we consider two samples of gas mass fraction of galaxy clusters obtained from X-ray surface brightness observations and two sub-samples of the SNe Ia taken from the Union2 compilation (Amanullah {\it{et al.}} 2010).

\subsubsection{La Roque et al. (2006)}

This sample comprises 38 massive galaxy clusters lying in the redshift range $0.14 < z < 0.89$, which were obtained from Chandra X-ray and OVRO/BIMA interferometric Sunyaev-Zel'dovich effect (SZE) measurements (La Roque {\it{et al.}} 2006). In order to perform a realistic model for the cluster gas distribution and take into account the possible presence of cooling flow, the gas density was modeled  by the non-isothermal double $\beta$-model, which generalizes the single $\beta$-model profile, introduced by Cavaliere and Fusco-Fermiano (1976) and the double $\beta$ model proposed by Mohr {\it{et al.}} (1999). Therefore, the cluster plasma and dark matter distributions were analyzed assuming hydrostatic equilibrium model and spherical symmetry, thereby accounting for radial variations in density, temperature and abundance.

\subsubsection{Ettori et al. (2009)}

This sample also comprises 38 galaxy clusters lying in the redshift interval $0.057 < z < 0.734$. In reality, this is a subsample of the galaxy cluster sample contenting 57 objects compiled by Ettori {\it{et al.}} (2009). To estimate the gas and total mass profiles, the electron density and temperature profiles were described by a functional form adapted from Vikhlinin {\it{et al.}} (2005). Finally, it is worth mentioning that, although La Roque {\it{et al.}} (2006) and Ettori {\it{et al.}} (2009) used a spherical model to describe the clusters, their temperature and eletronic profiles are different.  The fiducial model used in both samples was the $\Lambda${CDM} scenario whose angular diameter distance is given by
\bqn
\lb{ADDLCDM}
D^{*}_A = \frac{c H_0^{-1}}{1+z} \int^{z}_{0}{\frac{dz'}{\sqrt{\Omega_{M,0} (1+z')^{3} + \Omega_{\Lambda,0}}}} \; {\rm{Mpc}}\;,
\eqn
with $\Omega_{\rm{M,0}} = 0.3$, $\Omega_{\rm{\Lambda,0}} = 0.7$ and $H_{\rm{0}} = 70 {\rm{km \cdot s^{-1} \cdot Mpc^{-1}}}$. % and $c$ is the speed of light.

\subsection{SNe Ia subsample}

We built two sub-samples of $D_L$ measurements obtained from the Union 2 SNe Ia compilation (Amanullah {\it{et al.}} 2010) using the following criterion: for each galaxy cluster in the La Roque et al. and Ettori et al. samples we seek for one SNe Ia  whose redshift matches  that of  the respective galaxy cluster ($\Delta z<0.006$), thereby allowing a direct test of the CDDR. We end up with two pairs of sub-samples, both containing 38 galaxy clusters and SN Ia. The $D_L$ values are obtained from

\begin{equation}
 D_L = 10^{(\mu - 25)/5} {\rm{Mpc}}\;,
\end{equation}
where $\mu$ is the distance modulus, which does not depend on the validity of the CDDR. Figure 1a shows the redshift difference between the SNe Ia of our sub-samples and the galaxy clusters of the La Roque {\it{et al.}} (2006) (filled blue circles) and the Ettori {\it{et al.}} (2009) (filled red squares) samples. The largest difference in redshift is $\Delta z \leq 0.006$ and $\Delta z \leq 0.001$ for the La Roque {\it{et al.}} and Ettori {\it{et al.}} (2009) samples, respectively.

% \begin{figure}
% \label{Fig}
% \centerline{\epsfig{figure=fig2.eps,width=3.2truein,height=3.0truein}
% \hskip 0.1in}
% \caption{ The likelihood distribution functions for La Roque {\it{et al.}} (solid blue line) and Etorri {\it{et al.}} (red dashed line) samples using the linear parametrization P1.}
% \end{figure}
%
% \begin{figure}
% \label{Fig}
% \centerline{\epsfig{figure=fig3.eps,width=3.2truein,height=3.0truein}
% \hskip 0.1in}
% \caption{The same as in Figure 2 using the nonlinear parametrization P2. Comparing these results with those displayed in the previous figure, we clearly see that the $\eta$ parameterizations adopted in our analysis do not alter considerably the best-fit results. }
% \end{figure}

\section{Analysis and results}

Following Holanda {\it{et al.}} (2010), we parametrize a possible departure from the CDDR ($\eta = 1$) using two functional forms for $\eta(z)$, i.e.,
\begin{eqnarray}
\label{parameterizations}
\eta(z) = \; \left\{
\begin{tabular}{l}
$1 + \eta_{0} z$
\,
\quad \quad  \quad  \quad \quad \quad \quad \quad (P1)\\
\\
$ 1 + \eta_{0}z/(1+z)$
\quad \quad  \quad \quad \quad (P2) \\
\end{tabular}
\right.
%\nonumber
\end{eqnarray}
The first expression is a continuous and smooth linear one-parameter expansion  whereas the second one includes a possible epoch dependent correction which avoids the divergence at high-$z$.

By combining Eqs. (1) (with $\eta \neq 1$) and (\ref{GasFrac4}), we define
\begin{equation}
\label{etateste}
\eta_{obs}(z)=\frac{D_L^3f_{gas}^2}{N^2(1+z)^6D^{*3}_{A}}\;.
\end{equation} \label{chi2}
The likelihood  estimator is determined by $\chi^{2}$ statistics
\begin{equation}
\label{chi2} \chi^{2} = \sum_{z}\frac{{\left[\eta(z) - \eta_{obs}(z)
\right] }^{2}}{\sigma^2_{\eta_{obs}} }\;,
\end{equation}
where $\sigma^2_{\eta_{obs}}$ takes into account the propagation of the statistical errors in Eq. (\ref{etateste}). In our analyses, the normalization factor $N$ [see Eq. (\ref{GasFrac})] is taken as a nuisance parameter so that we marginalize over it. Since La Roque {\it{et al.}} (2006) sample presents assymmetric error bars, the data were treated using the D'Agostini (2004) method, i.e., $f_{gas} =  \widetilde{f}_{gas} + \Delta_+ - \Delta_-$, with $\sigma_{f_{gas}} = (\Delta_+ +\Delta_-)/2$, where  $\widetilde{f}_{gas}$ stands for the La Roque et al. (2006) measurements and $\Delta_+$ and $\Delta_-$ are, respectively, the associated upper and lower errors bars. Moreover, given that the largest difference in redshift is of $\leq 0.006$ for La Roque {\it{et al.}} sample ($\leq 0.001$ for  Ettori {\it{et al.}} sample), our results are not modified if we choose to use $z_{\rm{cluster}}$ or $z_{\rm{SNe}}$ in Eq. (\ref{etateste}).

 We show the likelihood distribution as a function of  $\eta_0$ for P1 (Fig. 1b) and P2 (Fig. 1c) by considering the La Roque {\it{et al.}} (blue solid line) and Ettori {\it{et al.}} samples (red dashed line). Regardless of the CDDR parameterization adopted, we clearly see that the La Roque {\it{et al.}} (2006) plus SNe Ia sample is in perfect agreement with the $\eta_0 = 0$ value ($\eta = 1$) whereas the Ettori {\it{et al.}} plus SNe Ia data presents a significant conflict. In particular, this latter combination of data provides $\eta_0 = -1.60^{+0.90}_{-0.70}$ (2$\sigma$) for the non-linear parameterization (P2), which is $\simeq 3.5\sigma$ off from the CDDR value $\eta_0 = 0$. { In Figures 2a and 2b we show $\eta_{obs}$  as a function of $z$ and the best-fit curves for the $\eta(z)$ parametrization P1 and P2.}

{ For the sake of completeness, we repeated the analysis taking into account a possible redshift dependence of the depletion parameter, as given by Ettori et al. (2008), i.e.,  $b(z)=0.923(\pm 0.006) + 0.032(\pm 0.01)z $, and marginalizing over the other quantities that compose the normalization parameter $N$. We found that the results are in full agreement with those derived previously (marginalizing over $N$).} Moreover, as argumented by La Roque et al. 2006, when marginalizing over $N$,  the systematic uncertainty should be negligible compared to statistical uncertainty, since most of the systematic uncertainties affect the normalization and do not introduce significant trends with redshift. It is also worth observing that for all analyses performed in this paper, a negative value of $\eta_0$ is prefered. In principle, such a result can be explained in terms of cosmic opacity or the existence of axion-like and mini-charged particles (see, e.g., Avgoustidis {\it{et al.}} (2010) and Jaeckel and Ringwald (2010) for a recent review on this subject). We summarize the main results of our analyses in Table I.

\begin{figure*}
\label{Fig}
\centerline{\epsfig{figure=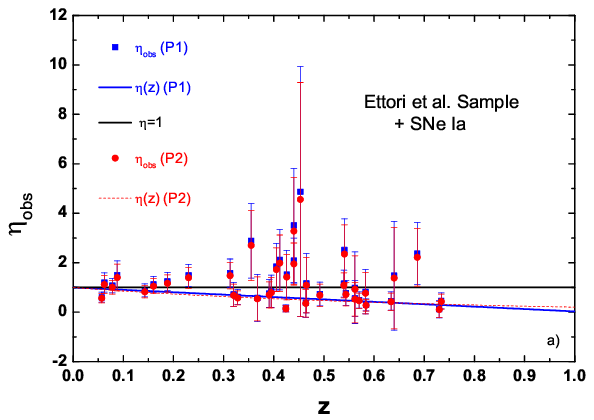,width=3truein,height=2.8truein}
\hspace{0.5cm}
\epsfig{figure=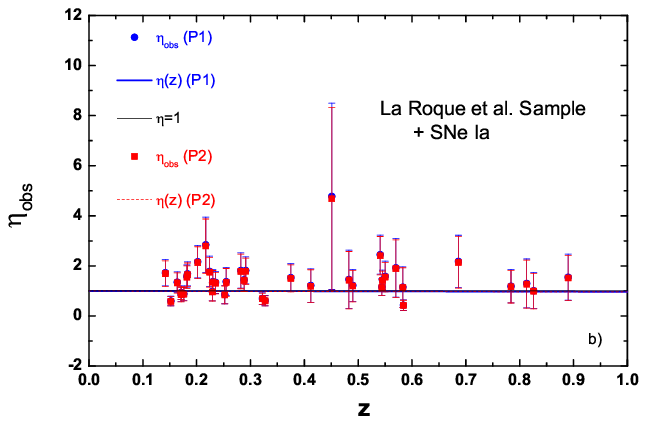,width=3.0truein,height=2.8truein}
\hskip 0.1in}
\vspace{-0.3cm}
\caption{ $\eta_{obs}$ as a function of the redshift. The curves stand for the best-fit values of $\eta (z)$ for both parametrizations P1 and P2, as given in Table I. Note that the P1 and P2 curves for La Roque  et al. behave as $\eta(z) \simeq 1$.}
\end{figure*}

\begin{table}
\caption{{Constraints on the CCDR.}}
\begin{center}
\begin{tabular} {|c|c|c|}
\hline\hline \\ \quad \quad  La Roque et al. plus SNe Ia sample \quad \quad  & \quad \quad $\chi^{2}/d.o.f$ \quad \quad
\\ \\ \hline \hline
$\eta_{0}$ (P1) = $ -0.03^{+1.03}_{-0.65}$ (2$\sigma$)& 49.61/37    \\
$\eta_{0}$ (P2) = $-0.08^{+2.28}_{-1.22}$ (2$\sigma$)&49.60/37 \\
\hline\hline  \\ Ettori et al. plus SNe Ia sample &  \quad \quad $\chi^{2}/d.o.f$  \quad \quad
\\ \\ \hline \hline
$\eta_{0}$ (P1) = $-0.97^{+0.54}_{-0.38}$ (2$\sigma$) & 41.78/37    \\
$\eta_{0}$ (P2) = $-1.60^{+0.90}_{-0.70}$ (2$\sigma$)& 41.65/37  \\
\hline \hline
\end{tabular} \label{table1}
\end{center}
\end{table}

\section{Conclusion}

In this letter, we have proposed a new cosmological model-independent test to the cosmic distance duality relation. We have discussed how measurements of the gas mass fraction of galaxy clusters together with SNe Ia observations can be used to impose limits on the $\eta$ parameter. We have also shown that, differently from most of the analyses discussed in the literatute, the $f_{gas}$ model function depends explictly on the CDDR [Eq. (\ref{GasFrac})].

To perfom our test we have used 38 gas mass fraction measurements of galaxy clusters recently studied by two groups considering different assumptions to describe the galaxy clusters (La Roque {\it{et al.}} and Ettori {\it{et al.}} samples) and two subsamples of 38 SNe Ia extracted from the Union2 compilation with $\Delta z \leq 0.006$. In order to take into account a possible influence of different $\eta$ parameterizations on the results we have used two different functions given by Eq. (14). We have shown that while the data set involving the La Roque {\it{et al.}} sample plus SNe Ia observations is in full agreement with the CDDR ($\eta = 1$), the sample built with Ettori {\it{et al.}} clusters and SNe Ia measurements presents a significant conflict ($\simeq 3.5\sigma$).{  It is worth mentioning that La Roque {\it{et al.}} (2006) and Ettori {\it{et al.}} (2009) used two different values of $R_\Delta$ to evaluate the total and gas masses of galaxy clusters, $R_{{\Delta}_{2500}}$ and $R_{{\Delta}_{500}}$, respectively, where $R_\Delta$ describes the sphere within which the cluster overdensity with respect to the critical density is $\Delta$. As is widely known, gas mass fraction measurements are affected by $R_\Delta$ choice  and this fact plays an important source of systematic error in our analysis, since $\eta_{obs} \propto f^2_{gas}$. However, if the results of Ettori et al. sample are confirmed by other analyses for different $R_\Delta$ values, it would bring to light new evidence for new physics, such as photon coupling with particles beyond  the standard model of particle physics, variation of fundamental constants, absorption by dust, etc. (see, e.g., Avgoustidis {\it{et al.}} (2010) and references therein for a discussion). Our results, therefore, reinforce the interest in searching for new and independent methods to test the CDDR.}

\section*{Acknowledgments}

The authors thank CNPq and CAPES for the grants under which this work was carried out.


\begin{thebibliography}{99}

\bibitem{allen} Allen, S.W.,  Schmidt,  R.W. \& Fabian, A.C., 2002, Mon. Not. R. Astron. Soc. 334, L1.
\bibitem{allen2} Allen, S.~W., Schmidt, R.~W.,  Ebeling, H., Fabian, A.~C. and van Speybroeck, L., 2004, Mon.\ Not.\ Roy.\ Astron.\ Soc.\  {353}, 457.
\bibitem{Allen:2007ue} Allen, S. W., Rapetti, D. A., Schmidt, R. W., Ebeling, H., Morris, G. and Fabian, A. C., 2008, Mon.\ Not.\ Roy.\ Astron.\ Soc.\  {383}, 879.
\bibitem{union2} Amanullah, R. et al., 2010, ApJ, 716, 712
\bibitem{lverde} Avgoustidis, A., Burrage, C., Redondo, J., Verde, L., \& Jimenez, R., 2010, JCAP, 1010, 024
\bibitem{bk04} Basset, B. A., Kunz, M.,\ 2004, PRD, 69, 101305. %[astro-ph/0312443v2]
\bibitem{caval} Cavaliere, A. \& Fusco-Fermiano, R.,\ 1978, A\&A., 667, 70
\bibitem{daly} Daly, R. A. \& Djorgovski, S. G., 2003, ApJ, 597, 9
\bibitem{bem06} De Bernardis, F., Giusarma, E. \& Melchiorri, A.,\ 2006, Int. J. Mod Phys. D, 15, 759
\bibitem{dagostini} D'Agostini, G.,\ 2004 [physics/0403086]
\bibitem{ellis71} Ellis, G. F. R.\ 1971, ``Relativistic Cosmology'',  Proc.\ Int.\ School Phys.\ Enrico Fermi, R. K.\ Sachs (ed.),  pp. 104-182 (Academic Press: New York); reprinted in  Gen.\    Rel.\ Grav., 41, 581, 2009
\bibitem{ellis07} Ellis, G. F. R.,\ 2007, Gen.\ Rel.\ Grav., 39, 1047
\bibitem{eth33} Etherington, I. M. H.,\ 1933, Phil.\ Mag., 15, 761; reprinted in Gen.\ Rel.\ Grav., 39, 1055, 2007
\bibitem{ettori} Ettori, S. et al., 2009, A\&A, 501, 61
\bibitem{G99} Gurvitz, L. I., Kellerman, K.I. \& Frey, S.,\ 1999, A\&A,  342, 378
\bibitem{Holandaa}Holanda, R.~F.~L., Lima, J.~A.~S. \& Ribeiro, M.~B.,\ 2010, ApJL, 722, L233
\bibitem{Holanda11} Holanda, R.\ F.\ L.,  Lima,  J.\ A.\ S.\ \&  Ribeiro, M. B., 2011a
\bibitem{Holandab}Holanda, R.~F.~L., Lima, J.~A.~S. \& Ribeiro, M.~B.,\ 2011b, A\&A Letters, 528, L14
\bibitem{Jaeckel10} Jaeckel, J. and Ringwald, A., 2010,  Ann.\ Rev.\ Nucl.\ Part.\ Sci.\  {60}, 405.
\bibitem{Khedekar11} Khedekar, S. \& Chakraborti, S., 2011, PRL, 106, 22
\bibitem{lunion} La Roque, S. J. et al.,\ 2006, ApJ, 652, 917
\bibitem{Li2011} Li, Z., Wu, P. \& Yu, H., 2011, 729, L14
\bibitem{lima} Lima,  J.~A.~S.~, Cunha, J.~V.~ \&  Alcaniz, J.~S.~, 2003,  Phys.\ Rev.\  D, {68}, 023510
\bibitem{Mohr} Mohr, J. J., Mathiesen, B., Evrard, A. E. 1999, ApJ, 517, 627
\bibitem{djain} Nair, R., Jhingan, S. \&  Jain, D.,\ 2011 [arXiv:1102.1065]
\bibitem{Peebles} Peebles, P. J. E.\ 1993, ``Principles of Physical Cosmology'', (Princeton University Press: New Jersey)
\bibitem{Reese02} Reese, E. D. et al.\ 2002, ApJ, 581, 53
\bibitem{Sarazin} Sarazin, C. L. \ 1988, ``X-Ray Emission from Cluster of Galaxies'', (Cambridge University Press: Cambridge)
\bibitem{Sasaki96} Sasaki, S., 1996, PASJ, 48, L119
\bibitem{Schneider} Schneider, P., Ehlers, J. and Falco, E.~E.\ 1992, ``Gravitational Lenses'', XIV, 560 pp.~112, Springer-Verlag Berlin Heidelberg (New York).
\bibitem{lverde2} Stern, D., Jimenez, R., Verde, L., Kamionkowski M. \& Stanford, S.A., 2010, JCAP, 02,008
\bibitem{SunZel72} Sunyaev, R. A. \& Zel'dovich, Ya.B., 1972, Comments  Astrophys.\ Space Phys., 4, 173
\bibitem{tolman} Tolman, R. C., 1930, Proc. Natl. Acad. Sci. U.S.A. {{16}}, 511.
\bibitem{uzan} Uzan, J. P., Aghanim, N. \& Mellier, Y., 2004, Phys.\ Rev.\ D, 70, 083533
\bibitem{vik}	Vikhlinin, A. et al., 2006, ApJ, 640,  691
\end{thebibliography}
\end{document}